\documentclass[aps,twocolumn,showpacs,superscriptaddress,prl,10pt]{revtex4-1}
\usepackage{graphicx}
\usepackage{dcolumn}
\usepackage{bm}
\usepackage{color}
\usepackage{amsmath}
\usepackage{amsfonts}
\usepackage{subfigure}
\usepackage[makeroom]{cancel}

\newcommand{\COMMENTED}[1]{}

\begin{document}

\author{Ettore Vitali}
\affiliation{Department of Physics, The College of William and Mary, Williamsburg, Virginia 23187}

\author{Hao Shi}
\affiliation{Department of Physics, The College of William and Mary, Williamsburg, Virginia 23187}

\author{Mingpu Qin}
\affiliation{Department of Physics, The College of William and Mary, Williamsburg, Virginia 23187}

\author{Shiwei Zhang}
\affiliation{Department of Physics, The College of William and Mary, Williamsburg, Virginia 23187}

\title{Visualizing the BEC-BCS crossover in the two-dimensional Fermi gas: \\
pairing gaps and dynamical response functions from \emph{ab initio} computations}

\begin{abstract}
Experiments with ultracold atoms provide a highly controllable laboratory setting
with many unique opportunities for precision exploration of quantum many-body 
phenomena. 
The nature of such systems, with strong interaction and quantum entanglement,  
makes 
reliable theoretical calculations challenging. 
Especially difficult are
excitation and dynamical properties, which
are often the most directly relevant to experiment.
We carry out exact numerical calculations, by Monte Carlo sampling of imaginary-time propagation 
of Slater determinants, to 
compute  the pairing gap 
in the two-dimensional Fermi gas
from first principles.
Applying state-of-art analytic continuation
techniques, we obtain 
the spectral function, and the density and spin structure factors providing unique tools
 to visualize the BEC-BCS crossover.
These quantities will allow
for a direct comparison with experiments.  
\end{abstract}

\maketitle

It is truly unusual when, 
starting from a microscopic Hamiltonian, theory can achieve
an exact description of a strongly correlated fermionic system
which, at the same time, can be realized in a laboratory
with great precision and control.
Experiments with ultracold atoms \cite{RevModPhysZwerger,RevModPhysStringari} have 
provided a possibility to realize such a scenario.
The accuracy that
can be reached in experiments with Fermi atomic gases and optical lattices is exceptional, 
thus offering a unique setting to explore highly correlated quantum fermion systems.
In this paper, we demonstrate that,
from the theoretical side, advances in computational methods now make it feasible to obtain
numerically exact results for not only  equilibrium properties,
but also excited states. 
 We compute the
 pairing gap,
spectral functions and dynamical
response functions in the two-dimensional Fermi gas across the range of interactions, 
which will allow direct comparisons with
spectroscopy or scattering experiments.
The dynamical properties provide a powerful tool
to probe the behavior of the system and to 
visualize the crossover from a gas of molecules
to a BCS superfluid.

We study the Fermi gas with a zero-range attractive 
interaction, which has generated a great deal of 
research activity \cite{RevModPhysZwerger,RevModPhysStringari,PhysRevLett.91.050401,PhysRevLett.105.030404,PhysRevLett.106.105301,PhysRevLett.107.145304,PhysRevLett.109.020406, PhysRevLett.112.045301,PhysRevLett.114.110403,PhysRevLett.115.010401,PhysRevLett.114.230401,PhysRevLett.116.045302,PhysRevLett.116.045303,PhysRevLett.95.060401,PhysRevA.93.033604}. The interest of the system is very wide, ranging from condensed
matter physics \cite{PhysRevLett.112.135302,PhysRevLett.107.145304} to nuclear physics, with possible important applications
also in the study of neutron stars \cite{ColdAtomsNeutronStars,PhysRevA.94.043614}. 
This system describes experiments with a collection of atoms, for example $^6$Li, which are cooled to
degeneracy in an equal mixture of two hyperfine ground states,
labeled $|\uparrow\rangle$ and $|\downarrow\rangle$.
Feshbach resonances allow the tuning of the interactions by varying an external magnetic
field, making the system a unique laboratory to explore many-body physics \cite{RevModPhys.82.1225,Bloch2012Quantum}.
Starting from a weakly interacting
BCS regime, where the attraction between particles
induces a pairing similar to the one observed in ordinary superconductors, 
a crossover is observed as  the interaction strength is
increased, 
leading
to a BEC regime where the Cooper pairs are tightly bound such 
that the system behaves as a gas of bosonic molecules.
While both the BCS and the BEC regimes are well understood,
the crossover regime
provides an excellent  
example of a strongly interacting quantum
many-body system \cite{RevModPhysZwerger,RevModPhysStringari}.

We focus in particular on the two-dimensional (2D) Fermi gas,
which has recently been realized experimentally using an highly anisotropic trapping potential \cite{PhysRevLett.105.030404,PhysRevLett.106.105301}.
The 2D system is important, since some of the most interesting   
physical phenomena,
such as high temperature superconductivity \cite{RevModPhys.78.17}, 
Dirac fermions in graphene \cite{RevModPhys.81.109} and topological superconductors \cite{RevModPhys.83.1057}, 
nuclear ``pasta'' phases \cite{PhysRevLett.90.161101} in neutron stars 
are two-dimensional in nature.
Quantum fluctuations are known to be enhanced in 2D, making it even more important 
to have quantitative results beyond mean-field approaches.

Experiments are just beginning to measure properties in the 2D gas  \cite{nature_Feld,PhysRevLett.112.045301,PhysRevLett.114.230401,PhysRevLett.116.045303,PhysRevA.94.031606}. An array of 
calculations have been performed \cite{PhysRevLett.112.135302,PhysRevA.92.023620,Klawunn20162650,Hao-2DFG,PhysRevA.93.023602,Giorgini}, although much less is available in 2D compared to three-dimensional 
systems.
The most direct connection with experiments is through response functions
and  accurate many-body data on dynamical response at low temperatures would
provide crucial and fundamental missing link.
However, these  are much more challenging theoretically and computationally \cite{Lianyi}.

In this paper, we develop the capabilities to 
 obtain unbiased results
for imaginary-time correlation functions in spin-balanced Fermi gas systems,
using first principles auxiliary-field quantum Monte Carlo (AFQMC) \cite{AFQMC-lecture-notes-2013,hubbard_benchmark,PhysRevB.94.085103,PhysRevB.88.125132} methods.
This provides a unique approach  
to excitations and
dynamical response functions. 
Focusing on the BEC-BCS crossover regime, we compute the
pairing gap as a function of the interaction strength, the
spectral function, which can be measured
experimentally in photoemission spectroscopy \cite{nature_Feld}, and the
density and spin structure factors, which can be measured
in two-photon scattering experiments \cite{PhysRevLett.109.050403}. 

As the range of the interaction in the Fermi gas system of cold atoms is much smaller than
the average inter-particle distance, the system can be modeled using a lattice 
Hamiltonian \cite{PhysRevA.86.013626}:
\begin{equation}
\label{hamiltonian}
\hat{H} = t\sum_{{\vec{k}},\sigma} \, \varepsilon({\vec{k}}) \, \hat{c}^{\dagger}_{{\vec{k}},\sigma}
\hat{c}^{}_{{\vec{k}},\sigma} + U \sum_{i} \hat{n}_{i,\uparrow}  \hat{n}_{i,\downarrow}\,,
\end{equation}
where the label $i$ runs over a square lattice
with $\mathcal{N}_s = L \times L$ sites hosting a total of $\mathcal{N}_p$ fermions, half with each spin
 $\sigma$ ($=\uparrow$
or $\downarrow$). 
The momentum ${\vec{k}}=(k_x,k_y)$
is defined on the reciprocal lattice with units $2\pi / L$
and 
$k_x$,$k_y\in[-\pi,\pi)$.
The dispersion is $\varepsilon({\bf{k}}) = k_x^2 + k_y^2$ and
$t = \hbar^2/(2 m b^2)$, with $b$ the lattice parameter.
The attractive on-site interaction $U/t$
is tuned \cite{PhysRevA.86.013626,Hao-2DFG} for each lattice density $n=\mathcal{N}_p/\mathcal{N}_s$
and Fermi momentum $k_F = \sqrt{2 \pi n} / b$ to produce the
desired scattering length $a$, defined as
the position of the node of the zero-energy $s$-wave solution
of the two-body problem. 

The ground state wave 
function $|\Psi_0 \, \rangle$ of $\hat{H}$  
is sampled using the AFQMC method  \cite{Hao-2DFG}.
For hamiltonian \eqref{hamiltonian} with $U/t < 0$, the sampling
is not affected by the sign problem, so that numerical results  can be obtained  free of any bias for 
each set of parameters $\{ \mathcal{N}_s, \mathcal{N}_p, U/t\}$.
Accelerated sampling techniques with force bias are used, together with other 
technical improvements \cite{Hao-inf-var}, which greatly improves the efficiency of our calculations. 
This allows us to reach large system sizes in order to reliably extrapolate to the continuum and 
then to the thermodynamic limit \cite{Hao-2DFG}.

Our computation of the dynamical correlation functions here relies on a new algorithm 
which improved the computational scaling in the calculation of 
imaginary-time correlation functions \cite{ettoreGAP} 
from $\mathcal{O}(\mathcal{N}_s^3)$ in standard
algorithms \cite{hirsch,assaad_prb,assaad_prl,jel_dyn_method,jel_dyn_method2,Alhassid_2014} to $\mathcal{O}(\mathcal{N}_s\,\mathcal{N}_p^2)$.
The algorithm lets fluctuations related to creation/destruction operators
or density/spin operators propagate in imaginary time, coupled to the stochastic 
evolution of the underlying AFQMC random walk or path-integral  \cite{ettoreGAP}. 
The dynamical correlation functions 
are obtained as suitable combinations
of matrix elements involving the Slater determinants  \cite{ettoreGAP}.
In the Fermi gas systems, the calculation is at the dilute limit, with 
$\mathcal{N}_s \gg \mathcal{N}_p $, so that a drastic speedup is achieved.
This allows us to study lattices of $\mathcal{N}_s \sim 2000$ sites in order to, as illustrated below, 
reach proper convergence of the results to the realistic limit.

The exact imaginary-time correlation functions allow one to access 
a number of important physical quantities.
We compute the pairing gap $\Delta$ 
from the large imaginary-time behavior of the dynamical Green's functions:
\begin{eqnarray}
G^p(\vec{k},\tau) &=& \left\langle \,\hat{c}^{}_{\vec{k}}
\,e^{-\tau ( \hat{H}-E_0)}\,\hat{c}^{\dagger}_{\vec{k}} \right\rangle 
\nonumber \\
G^h(\vec{k},\tau) &=& \left\langle \,\hat{c}^{\dagger}_{\vec{k}}
\,e^{-\tau ( \hat{H}-E_0)}\,\hat{c}^{}_{\vec{k}} \right\rangle\,,
\label{eq:Green}
\end{eqnarray}
where the superscripts $p$ and $h$ indicate particle and hole, and $E_0$ is the ground-state energy.
Moreover, we estimate the spectral function
\begin{equation}
A(\vec{k},\omega) = \left\langle \,\hat{c}^{}_{\vec{k}}
\,\delta( \omega - \hat{H})\,\hat{c}^{\dagger}_{\vec{k}} \right\rangle\,+\, \left\langle \hat{c}^{\dagger}_{\vec{k}}
\, \delta( \omega - \hat{H})\,\hat{c}^{}_{\vec{k}}\,\right \rangle
\label{eq:spectral-fcn}
\end{equation}
and 
the density and spin dynamical structure factors:
\begin{equation}
S^{\hat{O}}(\vec{k},\omega) = \left\langle \,\hat O_{\vec{k}}\,\delta( \omega - \hat{H})\,\hat O_{-\vec{k}}
\,\right \rangle\,,
\end{equation}
where the operator $\hat{O}$ is  
$\rho_{\vec{k}}=\hat{n}_{\vec{k},\uparrow} + \hat{n}_{\vec{k},\downarrow}$ for density 
and $S_{\vec{k}}=(\hat{n}_{\vec{k},\uparrow} - \hat{n}_{\vec{k},\downarrow})/2$ for spin, and
the brackets indicate  ground-state expectations.
These functions are obtained  from analytic continuation of the imaginary-time Green's functions and 
density-density or spin-spin correlation functions, using the
Genetic Inversion via Falsification of Theories (GIFT) method \cite{gift,giftREV,helium1D,rods1D,arrigoni_excitation_2013,he3_dyn,saccani_excitation_2012}.

\begin{figure}[ptb]
\begin{center}
\includegraphics[width=6.0cm, angle=270]{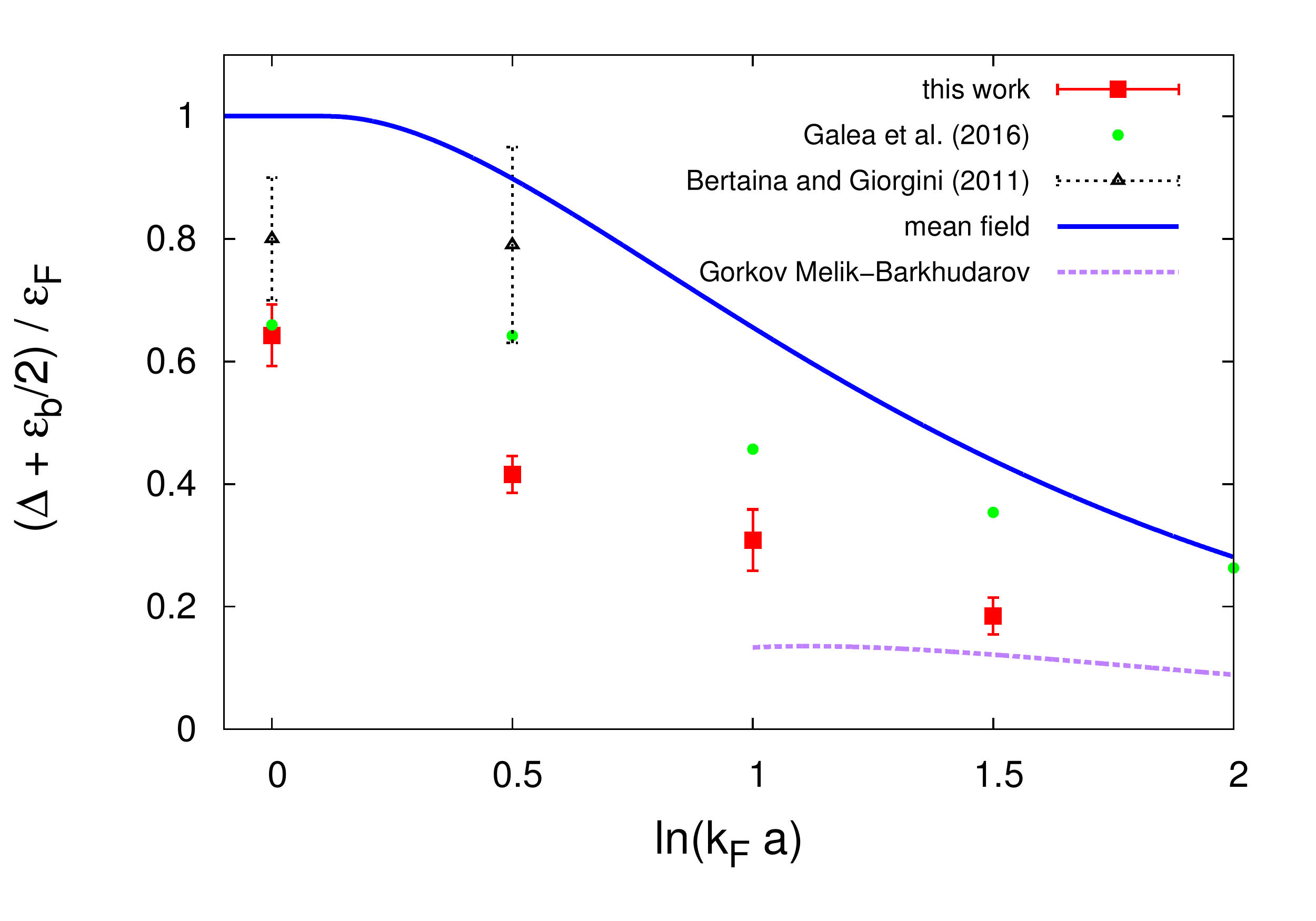}
\caption{ 
 (Color online) Pairing gap as a function of interacting strength, $\ln(k_F a)$. The 
gap values has been shifted by the binding energy, $\varepsilon_b$.
DMC results are from Refs.~\cite{PhysRevA.93.023602} (circles) and \cite{Giorgini} (triangles).
BCS mean-field result is also shown for reference.}
\label{fig:pairinggap}
\end{center}
\end{figure}

\begin{figure}[ptb]
\begin{center}
\includegraphics[width=7.0cm, angle=270]{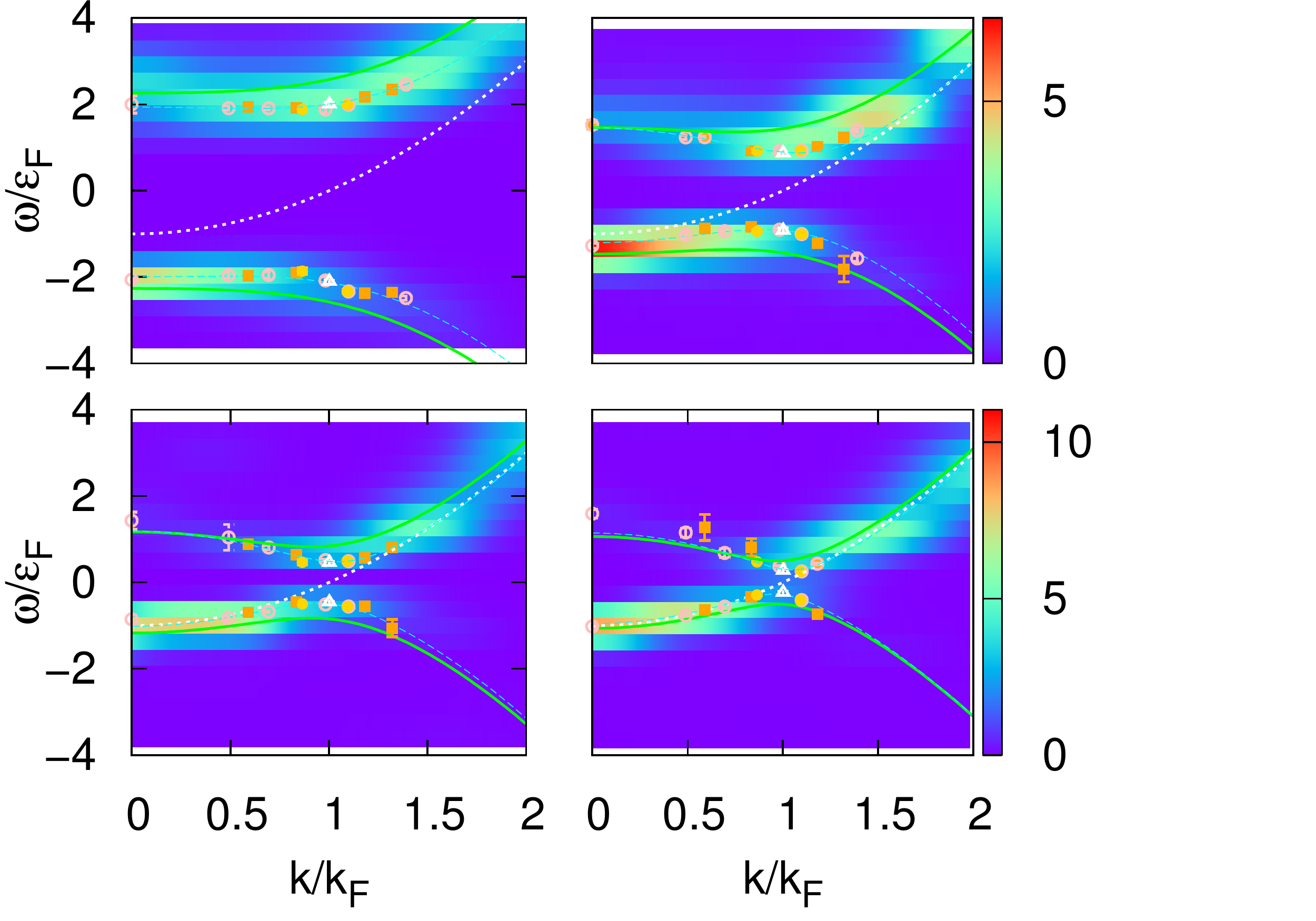}
\caption{ 
(Color online) Computed quasi-particle peaks and spectral functions. The four panels are for
different values of the interaction parameter: $\ln(k_F a) = 0$ (top left), $\ln(k_F a) = 0.5$ (top right),
 $\ln(k_F a) = 1$ (bottom left), $\ln(k_F a) = 1.5$ (bottom right).
Energies are measured in units of the Fermi energy $\varepsilon_F = \hbar^2 k_F^2/2m$
and momenta in units of the Fermi momentum $k_F$. The zero of the energy is
set to the chemical potential. 
The BCS-theory predictions for the quasi-particles
energies $E_{\pm}(\vec{k})$ are shown by solid lines, while 
the non-interacting spectral function is given by the dotted line. 
The symbols are the quasi-particle
peaks directly computed by AFQMC at the given momentum, for systems of $18$ particles on a $25 \times 25$ lattice (orange filled squares), $26$ particles on a $35 \times 35$ lattice (pink empty circles), $42$ particles on a $39 \times 39$ lattice (gold filled circles)
and $50$ particles on a $41 \times 41$ lattice (empty triangles).
Error bars are shown but some are smaller than symbol size.
The light dashed lines are interpolations in the neighborhood of the minimum. 
The color plots give the computed spectral functions.}
\label{fig:spectra}
\end{center}
\end{figure}

\begin{figure}[ptb]
\begin{center}
\includegraphics[width=10.0cm, angle=0]{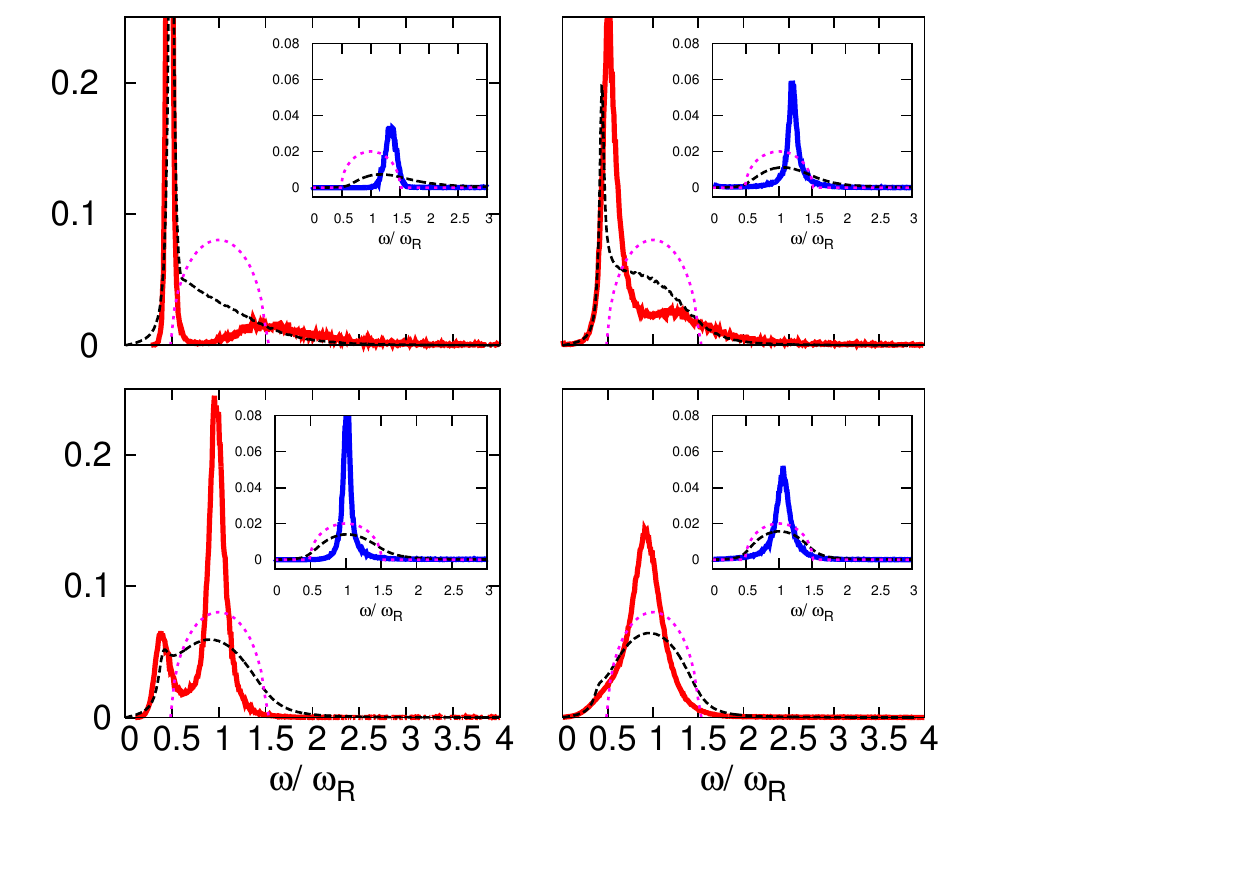}
\caption{
(Color online) Density (main graphs) and spin (insets) dynamical structure factors at $k = 4k_F$. 
The four panels show four different values of the interaction parameter: $\ln(k_F a) = 0$ (top left), $\ln(k_F a) = 0.5$ (top right),
 $\ln(k_F a) = 1$ (bottom left), $\ln(k_F a) = 1.5$ (bottom right).
Solid red lines are AFQMC results, while dashed black lines are obtained from dynamical BCS theory. The non-interacting
results are also shown (dotted magenta line) in each for reference.
The energies on the horizontal axes are measured in units of the atomic recoil energy,
$\omega_R = \hbar^2k^2/2m$.
}
\label{fig:sqomega}
\end{center}
\end{figure}

Figure~\ref{fig:pairinggap} shows the 
computed pairing gap across different interaction strengths.
The most standard approach to determine the pairing gap requires separate calculations corresponding to 
different particle numbers. While the spin-balanced calculation is free of the sign problem, the 
$({\mathcal{N}_p \pm 1})$ calculations are not. 
This makes our approach through imaginary-time Green's functions in Eq.~(\ref{eq:Green}) especially advantageous, 
since the Monte Carlo sampling 
remains at ${\mathcal{N}_p}$ and thus sign-problem-free  \cite{ettoreGAP}. We are able to 
determine the Green's function with high statistical accuracy 
in the asymptotic regime in imaginary time $\tau$,
so that the behavior is dominated by an exponential whose exponent gives the 
 quasi-particle peak at the targeted momentum  ${\vec k}$
(see Supplementary Materials).
A double exponential function is used in the fit to account for any residual effects.  
The imaginary-time interval on which the fit is performed is stochastically
varied and sampled,
and multiple data sets are generated to remove statistical correlations in imaginary time within each run.
 The final statistical uncertainty
 reflects the combined effects from the AFQMC error bars and the fitting procedure.
 We then scan ${\vec k}$
to locate the minimum/maximum (for the 
 particle/hole Green's functions) 
 for the pairing gap,
 as further illustrated in Fig.~\ref{fig:spectra}.

In Fig.~\ref{fig:pairinggap} we also show the BCS mean-field prediction, as well as
the current best many-body results, from 
recent diffusion Monte Carlo (DMC) calculations \cite{PhysRevA.93.023602,Giorgini}. 
It is seen that our pairing gap
is compatible with the DMC results on the BEC side 
of the crossover, but is consistently smaller
for larger values of $\ln(k_F a)$. The smaller gap value is
not surprising, since the DMC contains a fixed-node (FN) approximation which gives an upper 
bound on the computed energy. It is reasonable to expect that 
the trial wave function used for FN is of higher quality for the spin-balanced system compared 
than for the $(\mathcal{N}_p \pm 1)$ systems, which would lead to an 
overestimation of the pairing gap.
Our results on the BCS side are consistent with the rescaled BCS results
$\Delta_{BCS}/e$ from the theory by Gorkov and Melik-Barkhudarov, which
is expected to be exact in the BCS limit ($\log(k_Fa) >> 1$) \cite{Petrov, Gorkov}.

Figure~\ref{fig:spectra} plots the computed quasi-particle peaks as a function of $k\equiv |{\vec k}|$, 
together with the 
spectral function, for  four values of the interaction parameter.
The zero of the energy is
set equal to the chemical potential, which we can compute exactly \cite{Hao-2DFG}.
We will refer to the  function $A(\vec{k},\omega)$ as 
particle and hole spectral function  
respectively for $\omega > \mu$ and  $\omega < \mu$.
The particle spectral function originates from the first term on the right in Eq.~(\ref{eq:spectral-fcn}),
physically
representing states available for additional particles injected into the system,
while the hole spectral function, originating from the second term, contains information about states
occupied by the particles in the system, which are thus accessible by
the creation of holes.
In each panel, we show also mean-field
prediction for the quasi-particle
energies \cite{PhysRevLett.62.981}: $E_{\pm}(\vec{k}) = \pm\sqrt{\left( \hbar^2 k^2/2m - \mu_{\rm BCS}\right)^2 + \Delta_{\rm BCS}^2}$,
where $ \Delta_{\rm BCS}$ is the gap and $\mu_{\rm BCS}$ the chemical potential in BCS theory.
The non-interacting spectral function,
$A^0(\vec{k},\omega) = \delta \left( \omega - \left( \hbar^2 k^2/2m - \varepsilon_F\right) \right)$,
is also shown for reference.
In the AFQMC spectral functions obtained from the GIFT analysis, shown in the color plot,  quasi-particles
peaks are still visible, which are broadened from many-body correlations, resulting
in a non-zero imaginary part of the self-energy, and are 
renormalized with respect to the BCS dispersion relations.
The quasi-particle peaks computed directly from AFQMC are shown by symbols.
These were obtained following the procedure described in 
Fig.~\ref{fig:pairinggap}. Results from different system sizes are shown, which 
indicate convergence to the bulk limit within numerical resolution.
(Separate calculations were also 
carried out to verify that these densities are indistinguishable from the  
dilute limit \cite{Hao-2DFG}.)

The behavior of the spectral function provides a clear visualization of the
BEC-BCS crossover.
In the BEC regime at $\ln(k_F a) = 0$, a large gap, of the
order of the energy needed to break a molecule, separates the two branches,
which are roughly momentum-independent for $k \leq k_F$.
A smooth evolution of the spectral function is observed. 
In the BCS regime at $\ln(k_F a) = 1.5$, it 
starts to resemble the non-interacting behavior, where 
a gap is still present at the Fermi momentum, as in conventional
superconductors. The intermediate values of the interaction show
a smooth crossover between the two regimes.
Viewed in the reverse direction, gradual and significant departures from the BCS results are seen as 
the interaction strength is increased.

We also compute two-body dynamical correlations in imaginary time, which
can again be obtained using our method
with computational cost linear in ${\mathcal N}_s$  \cite{ettoreGAP}. 
From these, we apply analytic continuation to obtain 
the density and spin dynamical structure 
factors, $S^{\rho}(\vec{k},\omega)$ and
$S^{S}(\vec{k},\omega)$, which can be measured experimentally
using two-photons Bragg spectroscopy \cite{PhysRevLett.109.050403}. 
In particular, the high momentum behavior is  
very interesting
as it provides a highly sensitive probe of the BEC-BCS crossover.
We focus our attention on $k = 4 k_F$, close to the value recently
 investigated experimentally in three-dimensions \cite{PhysRevLett.109.050403}. 

The results are plotted
in Fig.~\ref{fig:sqomega}  
as functions of
the frequency $\omega$ for the four
values of the interaction parameter.
In addition to AFQMC, we have also performed
self-consistent dynamical
BCS theory calculations for the same system, following  the 
approach in Ref.~\cite{PhysRevA.74.042717} which studied
 the three-dimensional Fermi gas. 
 The results  are shown in the figure for comparison.
 Because the theory yields the response functions directly, it helps 
 to provide an additional gauge on the reliability of  analytic continuation analysis.
 We observe that the dynamical
BCS theory gives results on the response functions that are 
 qualitatively reasonable. 
 Significant differences arise from the AFQMC results, however,
 for example in the peak position in the spin structure factor for strong interactions.
 Direct comparisons of the  imaginary-time correlation functions show significant differences 
 between AFQMC and dynamical BCS theory as well,
  manifesting particle correlation effects absent in the latter.

In the density response, 
a large peak is seen at $\omega \simeq \omega_R/2$ 
in the deep BEC regime. Since 
 the particles are tightly paired to form molecules in this regime,
the response of the system at high momentum is dominated by the recoil of 
the molecules themselves, whose mass is twice 
the atomic mass. 
In contrast,   the response
on the BCS side is simply a free particle recoil with the bare mass of the atoms.
The behavior of the density response in the crossover regime
interpolates between the two physical pictures, as is evident
from Fig.~\ref{fig:sqomega}. 
Starting from $\ln(k_F a) = 0$, 
we observe
a gradual shift of the spectral weight from the dominant molecular contribution towards the second peak
at $\omega \simeq \omega_R$.
At $\ln(k_F a) = 1$, the second peak dominates, and the molecule peak almost disappears.
By $\ln(k_F a) = 1.5$, the response becomes qualitatively similar to the non-interacting one.

The spin response, on the other hand, is not sensitive to the
molecular mode at $\omega \simeq \omega_R/2$, since the positive
and negative fluctuations on the spin-$\uparrow$ and spin-$\downarrow$ particles
cancel each other. 
However, we observe that, as it happens in three-dimensions \cite{PhysRevLett.109.050403},
the intensity of the peak is smaller on the BEC side of the crossover,
and the position of the peak is shifted towards higher energies. This
corresponds to a suppression of the spin susceptibility, related to
the increased energy required to remove atoms from the molecules.

In summary, we have performed {\it{ab initio}} calculations of the pairing gaps and 
 dynamical correlation functions for the two-dimensional interacting
 Fermi atomic gas. 
 Numerically exact AFQMC 
predictions are provided for the pairing gap. 
From unbiased imaginary-time correlation
functions computed by AFQMC for the many-body ground state, 
 the spectral function and
the density and spin dynamical structure factors are obtained, via analytic continuation,
across the BEC-BCS crossover.
Much larger system sizes are reached in our simulations by the development and implementation of several 
technical advances. Many internal validations and self-consistency checks are performed and 
careful error quantifications are carried out to maximize the robustness and reliability of the results.
The results will allow benchmarks of further 
theoretical and computational developments, and direct comparisons with experiments.
The exact pairing gaps will also be crucial as an input 
for formulating density-functional theory in 2D \cite{tddft_Bulgac,1367-2630-18-11-113044}.  
Excitations and dynamical correlation functions provide excellent
 tools for visualizing the BEC-BCS crossover.
 In interacting many-fermion systems in general, they connect directly with experimentally accessible measurements.
Our approach opens up many new possibilities for the computational studies of strongly interacting
fermionic cold atomic systems. 
It is hoped that the results presented here will also serve as an illustration of state-of-the-art
computational capabilities, and 
will stimulate additional theoretical and experimental activities.
The feedback from such activities will in turn spur
 further computations and additional developments. 

We thank J.~Carlson, A.~Gezerlis, and Lianyi He for helpful discussions.
This work was supported by 
NSF (Grant No. DMR-1409510). E.V.~and S.~Z. were also supported by the  
Simons Foundation. Computing was carried out at the Extreme Science and Engineering Discovery Environment (XSEDE),
which is supported by National Science Foundation grant number ACI-1053575, and the computational
facilities at William and Mary.

%


\end{document}


\author{Ettore Vitali}
\affiliation{Department of Physics, The College of William and Mary, Williamsburg, Virginia 23187}

\author{Hao Shi}
\affiliation{Department of Physics, The College of William and Mary, Williamsburg, Virginia 23187}

\author{Mingpu Qin}
\affiliation{Department of Physics, The College of William and Mary, Williamsburg, Virginia 23187}

\author{Shiwei Zhang}
\affiliation{Department of Physics, The College of William and Mary, Williamsburg, Virginia 23187}

\title{Supplemental Material \\ Visualizing the BEC-BCS crossover in the two-dimensional Fermi gas: 
pairing gaps and dynamical response functions from \emph{ab initio} computations} 

\maketitle 

\section{Calculation of the pairing gap}

We show here how we compute the pairing gap $\Delta$
starting from the dynamical Green functions in imaginary time:
\begin{eqnarray}
G^p(\vec{k},\tau) &=& \left\langle \,\hat{c}^{}_{\vec{k}}
\,e^{-\tau ( \hat{H}-E_0)}\,\hat{c}^{\dagger}_{\vec{k}} \right\rangle 
\nonumber \\
G^h(\vec{k},\tau) &=& \left\langle \,\hat{c}^{\dagger}_{\vec{k}}
\,e^{-\tau ( \hat{H}-E_0)}\,\hat{c}^{}_{\vec{k}} \right\rangle\,,
\label{eq:Green}
\end{eqnarray}
The usual definition, involving the ground
state energies for systems with $\mathcal{N}_p \pm 1$
particles:
\begin{equation}
\begin{split}
& E_{0}^{\mathcal{N}_p + 1} - E_{0}^{\mathcal{N}_p} = \mu + \Delta \\
& E_{0}^{\mathcal{N}_p - 1} - E_{0}^{\mathcal{N}_p} = -\mu + \Delta
\end{split}
\end{equation}
can be recast in terms of the large imaginary time behavior
of the dynamical Green functions:
\begin{equation}
\label{aux_mat:exp}
G^p(\vec{k},\tau) \simeq c^p(\vec{k}) e^{- \tau \omega^p(\vec{k})},
\quad G^h(\vec{k},\tau) \simeq c^h(\vec{k}) e^{- \tau \omega^h(\vec{k})}
\end{equation}
since, for example:
\begin{equation}
E_{0}^{\mathcal{N}_p + 1} - E_{0}^{\mathcal{N}_p} = \min_{\vec{k}} \omega^p(\vec{k})
\end{equation}
Equation \eqref{aux_mat:exp}, for a finite system, follows
from the exact identity:
\begin{equation}
\label{aux_mat:exact}
G^p(\vec{k},\tau) = \sum_{n} | \langle \Psi^{\mathcal{N}_p}_0
| \hat{c}^{}_{\vec{k}}  \Psi^{\mathcal{N}_p + 1}_n \rangle |^2
e^{- \tau ( E_{n}^{\mathcal{N}_p + 1} - E_{0}^{\mathcal{N}_p})} \quad ,
\end{equation}
with a similar one for the holes.

The key quantities to be computed are thus:
\begin{equation}
\omega^{p,h}(\vec{k}) - \mu = \lim_{\tau \to +\infty} \phi^{p,h}(\vec{k},\tau) 
\end{equation}
where we introduce the notation:
\begin{equation}
\phi^{p,h}(\vec{k},\tau) = - \frac{ \log(G^{p,h}(\vec{k},\tau))}{\tau} - \mu
\end{equation}

\begin{figure*}[ptb]
\begin{center}
\includegraphics[width=12cm,angle=270]{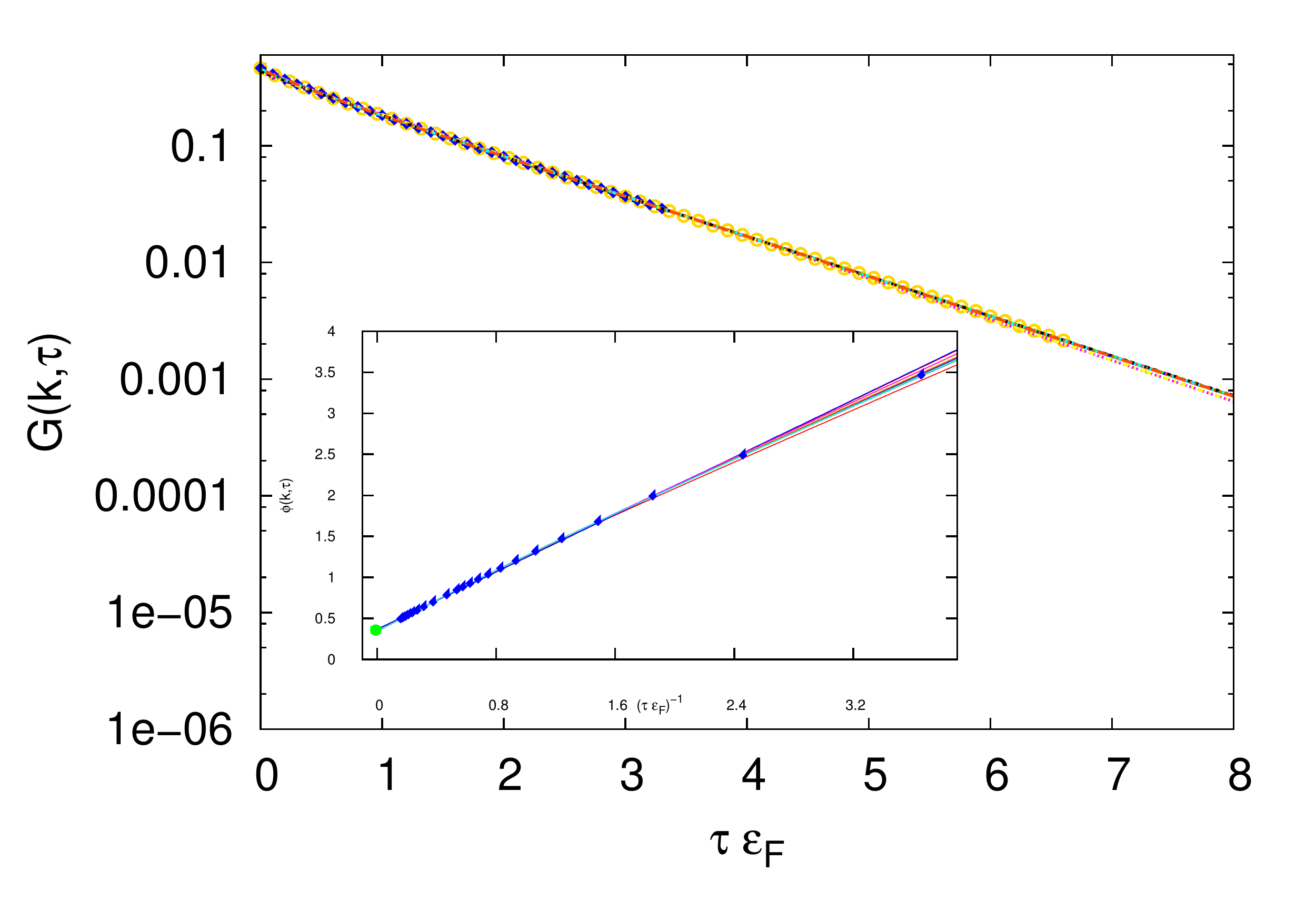}
\caption{(color online) $G^p(\vec{k},\tau)e^{-\tau \mu}$ for a system of $\mathcal{N}_p=26$ particles on
a $L \times L = 35 \times 35$ lattice, $|\vec{k}|/k_F = 0.98$.
The function $G^p(\vec{k},\tau)e^{-\tau \mu}$  has been computed as
an average over multiple independent AFQMC runs to avoid biases
due to autocorrelations in imaginary time.
Also shown are several double exponential fits.
The inset shows $\phi^{p}(\vec{k},\tau)$ in the large $\tau$ limit.
The double exponential fits are shown, also here. The circle represents
the AFQMC result with error-bar.
Error bars on $\phi^{p}(\vec{k},\tau)$ are smaller than the symbols size. }
\label{fig:big}
\end{center}
\end{figure*}

Since we can compute the
chemical potential exactly, we do not need both particle and
hole correlations functions. We checked, however, that
the two always give compatible results for the pairing gap.
From now on we will focus on the particles, for brevity.
\begin{equation}
\Delta = \min_{\vec{k}} \left( \omega^p(\vec{k}) - \mu \right)
\end{equation}
The procedure to obtain the value of the quasi-particle dispersion $\omega^p(\vec{k})$
from $G^p(\vec{k},\tau)$ is illustrated in the 
figures below.  Fig.~\ref{fig:small} shows results for a small system,
where exact diagonalization is available for comparison; 
Fig.~\ref{fig:big} shows results for a large system.

\begin{figure*}[ptb]
\begin{center}
\includegraphics[width=12cm,angle=270]{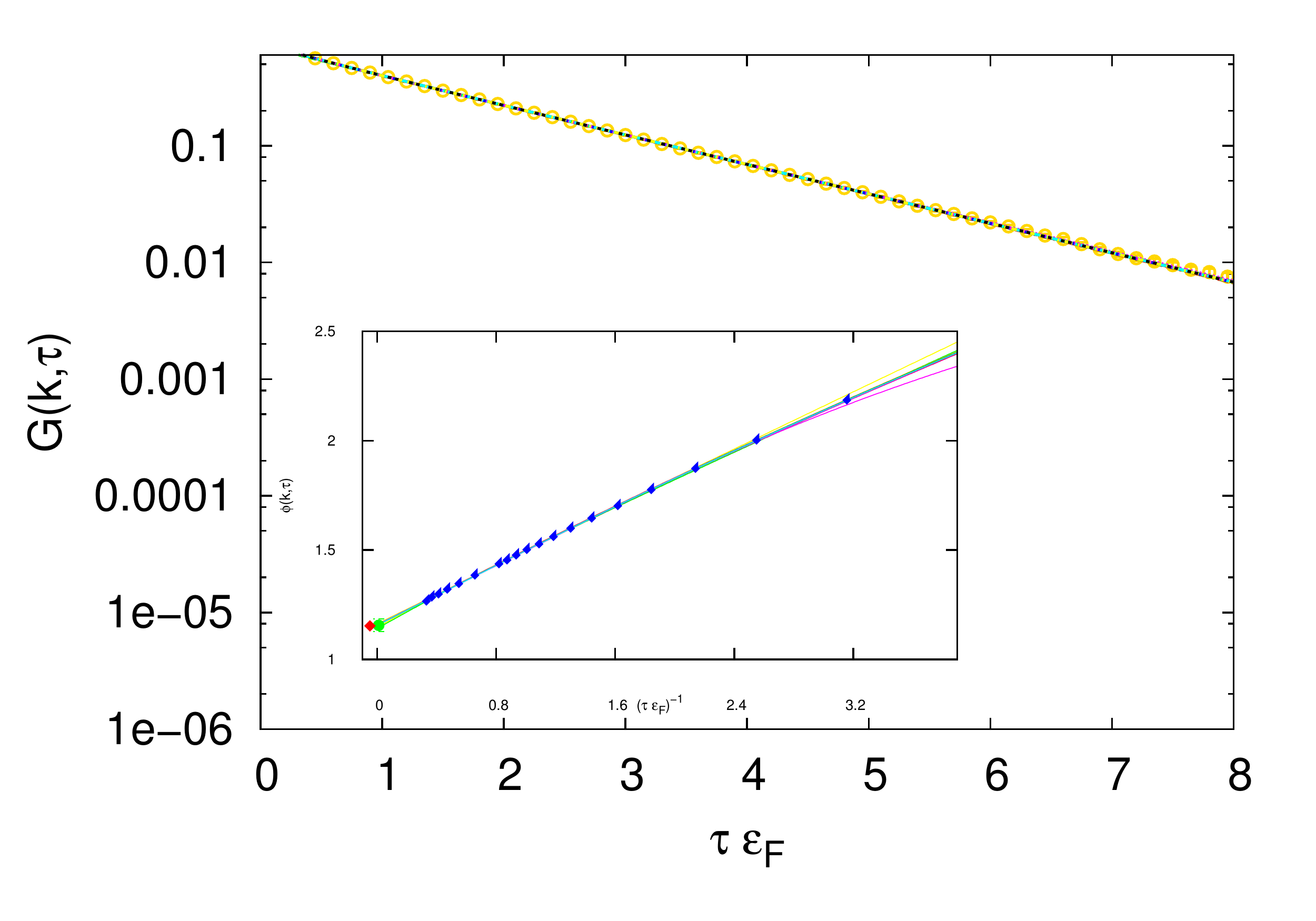}
\caption{(color online) $G^p(\vec{k},\tau) e^{-\tau \mu}$ from AFQMC for a system of $\mathcal{N}_p = 4$ particles on 
a $L \times L = 5 \times 5$ lattice, $\vec{k}= \frac{2\pi}{L}(1,0)$.
Also shown are several double exponential fits.
The inset shows $\phi^{p}(\vec{k},\tau)$ in the large $\tau$ limit. The
double exponential fits are also shown. The circle represents
the AFQMC result with error-bar ($\Delta/\varepsilon_F = 1.156 \pm 0.03$), while the diamond represents the
exact diagonalization result ($\Delta/\varepsilon_F = 1.153$).
Error bars on $\phi^{p}(\vec{k},\tau)$ are smaller than the symbols size.}
\label{fig:small}
\end{center}
\end{figure*}

The main plots show, in logarithm scale, the function $G^p(\vec{k},\tau) e^{-\tau \mu}$
for a given momentum, close to $k_F$. 
As mentioned in the main text, we average over independent simulations, in order to reduce the correlations
among data for different imaginary times.
We fit $G^p(\vec{k},\tau)$ with a linear combinations of two exponentials on an interval
$[\tau_0, \tau_{max}]$, the lower energy exponent yielding  $\omega^p(\vec{k})$, while
the higher energy exponential is meant to 
capture residual effects beyond \eqref{aux_mat:exp}.
The uncertainty on $\omega^p(\vec{k})$ comes from
a conservative combination of: (a) the AFQMC statistical error bars on $G^p(\vec{k},\tau)$, 
(b) uncertainty on the fitting parameters and (c) 
dependence on the choice of the interval
$[\tau_0, \tau_{max}]$, $\tau_0$ being randomly sampled
in the large imaginary time tail of $G^p(\vec{k},\tau)$. 

In the insets of the two plots we show the function
$\phi^p(\vec{k},\tau)$ in the large imaginary time
limit, together with the computed $\omega^p(\vec{k})$.
We plot several
examples of fitted functions, whose limits for $\tau \to \infty$ 
are used to compute $\omega^p(\vec{k})$ and the error bar. 
For the small system, we also show the exact diagonalization result, which is in excellent
agreement with the AFQMC result.

The consistency seen between the the small system in Fig.~\ref{fig:small}
and the large system in Fig.~\ref{fig:big} is a further indication of the robustness of
our calculations.

Finally, in table~\ref{aux-mat:tab} we list the values of
the computed $\omega^{p,h}(\vec{k})$
in a neighborhood
of the minimum, which will provide valuable benchmark.

\begin{table}[ptb]
\caption{Quasi-particles and quasi-holes dispersions}
\label{aux-mat:tab}
\begin{tabular}{lllll}
\hline
\hline
$\log(k_Fa)$ & $\varepsilon_B/2\varepsilon_F$ & $|\vec{k}|/k_F$ & $(\omega^p(\vec{k}) - \mu)/\varepsilon_F$  &  $(\omega^h(\vec{k}) + \mu)/\varepsilon_F$   \\
\hline
0 & -1.26 & 0.00  & 2.0(2)   & 2.06(6)    \\
0 & -1.26 & 0.49 & 1.91(5)  & 1.95(7)    \\
0 & -1.26 & 0.59 & 1.94(5)  & 1.96(5)    \\
0 &  -1.26 & 0.70 & 1.91(3)  & 1.95(7)    \\
0 &  -1.26 & 0.84 & 1.92(7)  & 1.90(7)    \\
0 &  -1.26 & 0.86 & 1.88(6)  & 1.86(8)    \\
0 &  -1.26 & 0.98 & 1.88(8)  & 2.08(8)    \\
0 &  -1.26 & 1.00 & 1.99(3)  & 2.12(5)    \\
0 &  -1.26 & 1.09 & 1.99(4)  & 2.33(4)    \\
0 &  -1.26 & 1.10 & 1.99(4)  & 2.33(4)    \\
\hline
0.5 & -0.46 & 0.84 & 0.93(6)  & 0.84(7)    \\
0.5 & -0.46 & 0.86 & 0.94(4)  & 0.92(4)    \\
0.5 & -0.46 & 0.98 & 0.92(5)  & 0.94(5)    \\
0.5 & -0.46 & 1.00 & 0.97(1)  & 0.87(1)    \\
0.5 & -0.46 & 1.09 & 1.00(5)  & 0.93(5)    \\
0.5 & -0.46 & 1.10 & 0.93(5)  & 1.00(5)    \\
0.5 & -0.46 & 1.18 & 1.02(7)  & 1.21(7)    \\
\hline
1.0 & -0.17 & 0.84 & 0.64(5)  & 0.45(2)    \\
1.0 & -0.17 & 0.86 & 0.47(6)  & 0.50(1)    \\
1.0 & -0.17 & 0.98 & 0.52(3)  & 0.51(1)    \\
1.0 & -0.17 & 1.00 & 0.45(2)  & 0.45(1)    \\
1.0 & -0.17 & 1.09 & 0.49(4)  & 0.55(1)    \\
\hline
1.5 & -0.06 & 0.86 & 0.47(1)  & 0.29(2)    \\
1.5 & -0.06 & 0.98 & 0.36(4)  & 0.26(3)    \\
1.5 & -0.06 & 1.00 & 0.27(2)  & 0.23(2)    \\
1.5 & -0.06 & 1.09 & 0.26(3)  & 0.27(3)    \\
\hline
\hline
\end{tabular}
\end{table}